# The Potential of MXene Materials as a Component in the Catalyst Layer for the Oxygen Evolution Reaction

Michelle P. Browne[a*], Daire Tyndall[a] and Valeria Nicolosi[ab]

[a]*Centre for Research on Adaptive Nanostructures and Nanodevices (CRANN), Advanced Materials and BioEngineering Research (AMBER) Centre, School of Chemistry, Trinity College Dublin, Ireland.*

[b]*I-Form Research Center, Trinity College Dublin, Dublin, Ireland*

*brownem6@tcd.ie

**Abstract**

MXenes are a class of 2D/layered materials which are highly conductive, hydrophilic, have a large electrochemical surface area and are easily processible into electrodes for energy applications. Since the discovery of MXenes over ten years ago, these materials have been mainly used in the preparation of electrodes for batteries and supercapacitors. However, due to their aforementioned properties, MXenes could potentially be utilised as a component in the catalyst layer for the Oxygen Evolution Reaction (OER). This opinion piece will discuss some of the recent literature in the area of hybrid catalysts consisting of various Transition Metal Oxides (TMOs) and MXenes for the OER. We will also discuss current drawbacks and future outlook in this new area of research.

## 1. Introduction

In an electrolyser device, the catalyst layer can contribute to losses in the overall running of the electrolyser due to inactive OER sites and low conductivity of the materials.[1,2] To achieve next generation inexpensive OER electrolyser catalysts, the catalysts themselves must be electrically conductive, mechanically and chemically stable under operating conditions, exhibit a high electrochemical surface area, and contain a high concentration of active sites for the evolution of $O_2$. This has not been achieved to date for Proton exchange Membrane (PEM) and Alkaline Anion Exchange Membrane (AAEM) water electrolysis. One avenue to explore to make a catalyst that possesses all these characteristics would be to essentially combine different materials that exhibit these properties individually and make a 'super' catalyst.

MXenes are a family of 2D materials, which are made up of transition metal carbides and nitrides, produced from MAX phases by various etching and delamination processes, Figure 1A. [3,4] A MAX phase has the general formula of $M_{n+1}AX_n$ where the M is an early transition metal, the A is an element from group 13 and 14 of the periodic table, and the X represents a carbon or nitrogen.[5] During the etching process, done in a fluoride ion based solution, the element from group 13/14 is removed from the MAX structure causing the carbide layers to become terminated by $OH^-$, $O^-$, $Cl^-$ or $F^-$ groups which are subsequently called 'surface groups or edge sites'.[5] The resulting structure is known as a 'MXene'.[5,6]

MXenes are known to be highly conductive, hydrophilic, and tuneable which are all advantageous properties that could lead to improving pure metal oxides when combined in a catalyst layer for OER. [6] As metal oxide materials lack high conductivity, which adds to the overall losses in an electrolytic cell, the addition of MXene materials could provide a high conductive support network making the hybrid material into a superior OER catalyst. Additionally, the hydrophilic nature of the MXenes will allow for the full coverage of $OH^-$ ions on the surface of the catalyst from the electrolyte which should help in the formation of $O_2$. Finally, being able to tune the various MXene materials could significantly improve the conductivity and the hydrophilic nature of the MXene, hence further improving the ability to evolve $O_2$. However, to date MXenes are not known to contain active sites for the OER, as no MXenes with metals for promoting the OER (e.g. Ni, Ru, Ir, Co, Mn or Fe) have been successfully synthesised (however MXenes containing Mn and Fe have been theoretically reported).[7,8] Interestingly, for the opposite water splitting reaction, the hydrogen evolution reaction (HER), MXene materials have been proven to be promising through computation calculations and experimentally methods.[9-11]

On the other hand, Transition Metal Oxides (TMOs) are an exciting group of materials, that possess various intriguing physical properties that can change depending on the oxidation state of the material and are known to be active OER catalysts, Figure 1B.[12,13] However, these TMO materials exhibit instability and dissolution during operation which renders these materials unsuitable for deployment in large-scale electrolyser devices.[14]

By combining inexpensive, active TMO catalysts with the conductive MXenes, most of the characteristics of the 'super' catalysts described previously, i.e. active site density due to the metal oxides and high conductivity due to the MXenes, in theory can be achieved.[15] The MXene materials may also provide a high surface area network for the TMO materials, similar to what has been attempted with carbon nanotubes (CNT) for OER composite materials, in order to improve electron transfer properties.[16] However, CNT based materials are known to contain metal impurities that enhance the OER (unaware to most) and corrodes under anodic OER potentials therefore are not the ideal composite materials with TMOs for the OER.[17,18] This high surface area network of a TMO/MXene composite will hopefully improve the operational stability of the catalyst layer due to improved mechanical properties.

Finally, combining MXenes and the TMOs may even result in a lower loading of the TMOs, further lowering the catalyst costs. Hence, the combination of TMOs and MXenes into heterostructured layers or functionalising the MXenes with TMOs is a new and exciting avenue to be explored for the generation of highly active OER catalysts in PEM and AAEM electrolysers.[19]

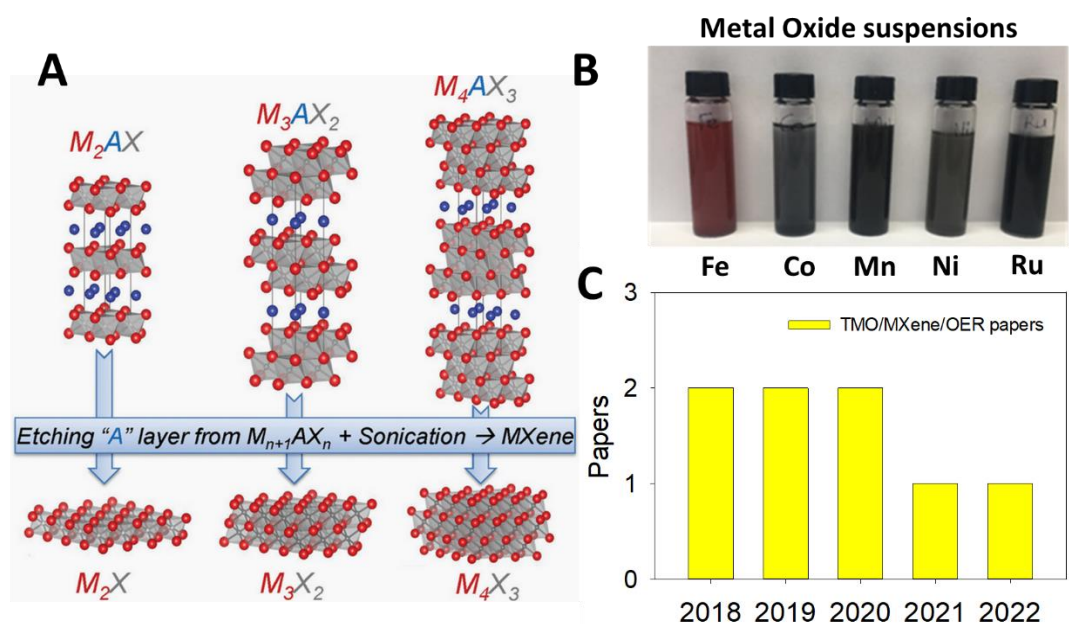

**Figure 1. A.** MXene fabrication (Reproduced with permission from Zhang *et al.* [20] Copyright Wiley (2014)) **B.** Metal Oxide Suspensions (Reproduced with permission from Browne *et al.* [21] Copyright Royal Society of Chemistry (2021)) **C.** Number of TMO/MXene papers for the OER (includes published literature from Scopus (01/02/2022) terms used 'Oxygen Evolution Reaction' 'MXene' and 'metal oxide'. One paper is a review and one paper is for batteries, not OER).

2. **TMO/MXene OER Catalysts in Literature**

To date only a hand full of papers have been published in this area, Figure 1C and Table 1. All of these papers have the same overarching conclusion that the addition of MXenes to transition metal oxides improves the initial OER performance compared to the TMO or the MXene alone.

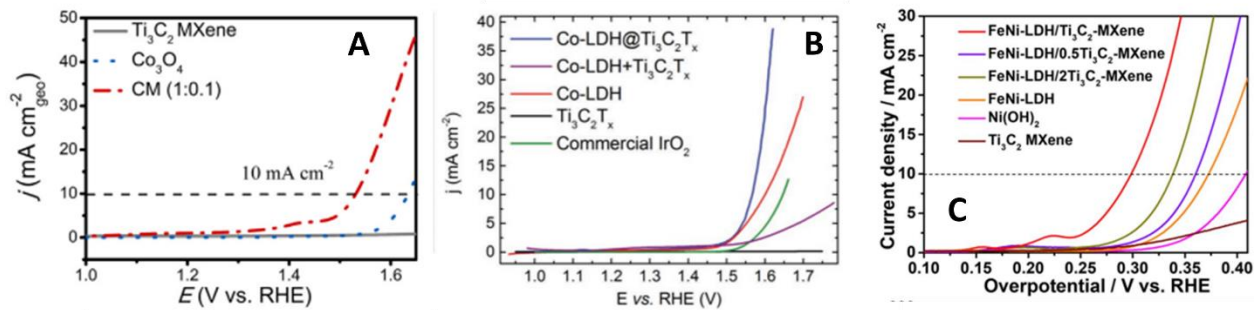

**Figure 2.** Linear Sweep Voltammetry Curves in the OER region showing hybrid TMO/MXene materials outperform their TMO and MXene counterparts from research published by **A.** Lu et al. (Reproduced with permission from Lu *et al.* [22] Copyright Elsevier (2020)) **B.** Benchakar et al. (Reproduced with permission from Benchakar *et al.* [23] Copyright Wiley (2019)) and **C.** Yu et al. (Reproduced with permission from Yu *et al.* [24] Copyright Elsevier (2018)).

For example, Lu *et al*. synthesised a hybrid MXene composite with $Co_3O_4$ decorated on $Ti_3C_2T_x$ MXene flakes by a solvothermal reaction at 150 °C for three hours.[22] The resulting $Co_3O_4$/ $Ti_3C_2T_x$ hybrid exhibited an OER overpotential of ~ 300 mV at a current density of 10 mA cm$^{-2}$ (the current benchmark used in literature when reporting the performance of OER materials) from linear sweep voltammetry measurements, Figure 2A.[22] Under the same OER conditions, the authors reported that a $Co_3O_4$ only material reached the same current density at overpotentials of 390 mV, while the $Ti_3C_2T_x$ can be deemed OER inactive as the current density at an overpotential of ~400 mV is virtually zero.[22] Furthermore, in this study, Lu and co-workers investigated the effect of the ratio of metal oxide:MXene on the OER. The four ratios of metal oxide:MXene prepared were 1:0.1, 1:0.4, 1:1 and 1:10. The results showed that the lowest amount of MXene to metal oxide (i.e. 1: 0.1) exhibited the best OER results in terms of overpotential at a current density at 10 mA cm$^{-2}$.

Benchakar and co-workers have observed a similar phenomenon for a Co layered double hydroxide (LDH)/ $Ti_3C_2T_x$ material fabricated by a polyol and solvothermal process.[23] In this particular study, the Co LDH/ $Ti_3C_2T_x$ outperformed the unsupported Co LDH for the OER by 50 mV, Figure 2B.[23] The authors have also reported that the MXene structure can be preserved for oxidation during synthesis and the OER by the well distributed Co LDHs on the surface of the MXene. Interestingly, the authors also reported that this Co LDH/ $Ti_3C_2T_x$ hybrid synthesized by chemical routes outperformed a material which consisted of the Co LDH and the $Ti_3C_2T_x$ mechanically mixed. This increase in performance of the chemically synthesized hybrid compared to the mechanically mixed catalyst was hypothesised to be due to the higher charge transfer resistance due to the close proximity of the Co LDHs and the $Ti_3C_2T_x$ or the lower amount of active sites in the mechanically mixed material.[23]

Additionally, the integration of MXene materials into a composite with bimetallic TMOs has been shown to be advantageous for the OER. Yu *et al.* reported that the synthesis of an FeNi-LDH/ $Ti_3C_2T_x$ material, fabricated by the co-precipitation of $Fe^{2+}$ and $Ni^{2+}$ from metal salts with already exfoliated MXene flakes under reflux, which also outperformed its individual counterparts under OER conditions, Figure 2C. The authors attributed the superior OER performance of the composite material to the increase in the charge transfer properties from electrochemical impedance measurements, a shift in the Ni redox peaks to more anodic potential that may induce the OER earlier and an enhancement in the O binding strength of the FeNi LDH due to an electron extraction as a result of the coupling with the MXene. [24]

| TMO | MXene | Synthesis method | Initial OER performance (Overpotential at 10 mA cm$^{-2}$) | Experimental reasoning for OER activity from postmortem or in-situ analysis | Ref |
|---|---|---|---|---|---|
| $NiFe_2O_4$ | $Ti_3C2T_x$ | Hydrothermal | 266 mV | None given | [25] |
| NiFe | $Ti_3C_2T_x$/ reduced graphene oxide | Metal oxides were precipitated and MXene/rGO mixed in | 235 mV | None given | [26] |
| Co LDH | $Ti_3C_2T_x$ | Polyol synthesis  Mixed | Polyol: 340  Mixed: Over 520 mV | None given | [23] |
| $Co_3O_4$ | $Ti_3C_2T_x$ | Solvothermal | 300 mV | None given | [22] |
| PtO-PdO NP | $Ti_3C_2T_x$ | Solution Plasma modification | 310 mV | None given | [27] |
| FeNi LDHs | $Ti_3C_2T_x$ | ionic hetero-assembly – metal salts and MXene refluxed | 298 mV | None given | [24] |

### 3. Current Drawbacks and Potential Avenues for Success (Outlook)

It is evident from literature that MXenes do significantly enhance the initial performance of TMO catalysts for the OER. However, MXene materials in a water-based solution are known to be unstable in air, Figure 3A.[28] The edge sites/surface groups of the MXene materials will oxidise first (due to deoxygenated species) to produce $TiO_2$ which will then lead to the whole flake becoming oxidized and hence decreasing the conductivity of the materials, Figure 3A. This can potentially be a huge problem for any electrochemical energy application including electrolysis.

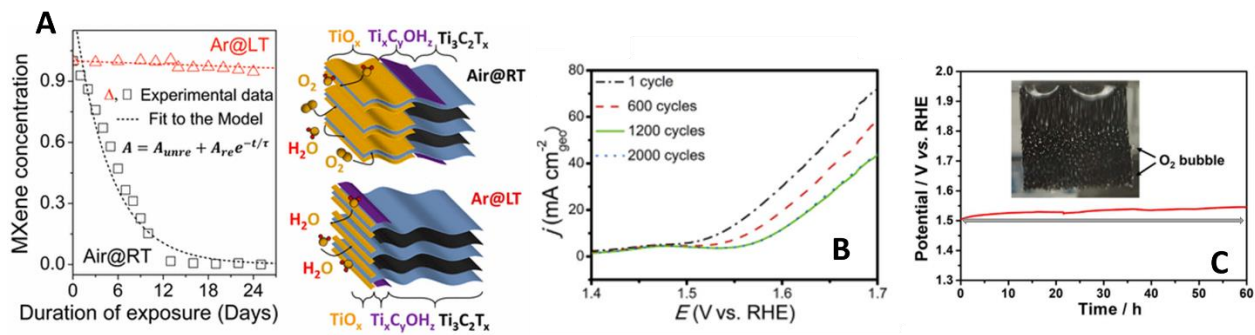

**Figure 3A.** Showing the decrease in MXene flake stability in air at room temperature (Reproduced with permission from Zhang *et al.*[28] Copyright American Chemical Society (2017)). **B.** Multi-cycling of a $Co_3O_4$/ $Ti_3C_2T_x$ hybrid material (Reproduced with permission from Lu *et al.*[22] Copyright Elsevier (2020)). **C.** Chronopotentiometry of a FeNi-LDH/ $Ti_3C_2T_x$ material on Ni foam (Reproduced with permission from Yu *et al.*[24] Copyright (2018)). Grey line with arrow was drawn by authors to show the increase in potential with time.

Unfortunately, the electrochemical instability of hybrid TMO/MXene materials have been already observed in the small amount of literature published in the area to date. For example, Lu *et al.* multi-cycled their $Co_3O_4$/ $Ti_3C_2T_x$ hybrid in the OER region for 2000 times and observed a significant decrease in activity over time, Figure 3B.[22] Furthermore, a FeNi-LDH/ $Ti_3C_2T_x$ material on Ni foam, synthesized by Yu and co-workers, also exhibited a decrease in activity over time during a chronopotentiometry test at 10 mA cm$^{-2}$ for 60 hours, Figure 3C.

The instability of these materials at such low current densities presents a major drawback in the potential for MXenes to be incorporated as a component of an OER catalyst layer. If these hybrid materials are to be employed in large scale electrolysis devices, the materials must remain stable at high current densities of 1 A cm$^{-2}$ and higher. In order to alleviate major instabilities in these hybrid materials, investigations using in-situ or operando measurements in conjunction with electrochemical techniques need to be carried out in order to determine the reasoning for the instability of the materials, which is likely to be related to the degradation of the MXene to $TiO_2$ under the extreme oxidative conditions present during the OER. The controlled synthesis of hybrid materials that can hinder the instability of the MXene component by covering the edge sites with active materials for the OER is another avenue which can be explored. Benchakar and co-workers have reported that their polyol/solvothermal synthesis method did fabricate a TMO/MXene hybrid material that is OER stable by covering the MXene edge sites with Co LDHs, however no stability tests were conducted to determine the long-term stability, which is needed to learn about possible TMO/MXene instabilities if superior such composite materials are to be designed in the future.[23]

A second route which could be undertaken to improve the OER performance and stability of TMO/MXene hybrids is to explore the (continuously expanding) world of MXenes materials. In literature the only MXene which has been utilised to synthesise TMO/MXene materials is $Ti_3C_2T_x$. There are numerous MXene materials now available and all exhibit different chemical and physical properties. [3,29,30] These other MXenes may be more stable and/or active for the OER when compared to the most common MXene, $Ti_3C_2T_x$. For example, it is well known that the incorporation/presence of Fe based materials/impurities into metal oxides improves the parent oxide towards the OER.[31] Therefore if the theoretically proposed $Fe_2CT_2$ MXene materials could be synthesized,[32] this may lead to an enhancement in the OER activity, while also attaining high conductivity and hydrophilicity of the combined TMO/MXene materials compared to when a non-Fe MXene (e.g. $Ti_3C_2T_x$) is present in the hybrid material.

A third route which could be undertaken to improve the performance of the metal oxide/MXene hybrid is to further investigate what ratio of TMO:MXene is optimum for the OER. Lu and co-workers showed that a ratio of 1:0.1 metal TMO:MXene is the best ratio for the OER in their study. However, this 1:0.1 TMO:MXene material contained the lowest ratio of MXene. Therefore, further studies into lower amounts of MXene in the hybrid need to be undertaken to determine if even less MXene is favourable for the OER when combined with metal oxides.

Furthermore, Gogotsi and co-workers have recently reported that MXene materials containing more than one metal can be fabricated with different stoichiometries and can all be tuned in respect to the chemical properties they exhibit.[3] If the chemical properties of the aforementioned mixed metal MXenes can be tuned, this opens up a huge space to investigate the suitability of these TMO/MXene hybrid materials for the OER.

Due to the large volume of TMO/MXene combinations, Density Functional Theory (DFT) will also play a vital role in the screening of the most promising TMO/MXene OER catalysts, as it has done for the HER.[11] DFT calculations, such as free energy and surface Pourbaix diagrams, could be utilised to predict stable surface terminations under OER conditions which could help experimental synthesis of stable TMO/MXene OER catalysts.

Finally, for these hybrid TMO/MXene catalysts to reach a stage of commercialisation, the TMO/MXene materials would need to outperform the current state-of-the-art materials in terms of activity and stability in actual electrolyser devices and not just in a conventional three-electrode cell used in OER studies throughout academia. The reason behind this, is that in a conventional three-electrode cell, it has been shown that OER catalysts do not behave the same as in an electrolyser device.[21,33,34] However, to reach a stage of testing TMO/MXene catalysts in an electrolyser device, first more fundamental issues must be tackled including the oxidation of MXenes under OER potentials, the optimum metal oxide to MXene ratio for OER and improving the overall activity of the hybrid catalysts by synthetic/ in-situ characterisation feedback mechanisms.

**Declaration of interest**

Nothing declared

**Acknowledgements**

M.P.B. would like to acknowledge the European Union's Horizon 2020 research and innovation programme under the Marie Skłodowska-Curie grant agreement No. 884318 (TriCat4Energy). V.N acknowledges the support of SFI AMBER (12/RC/2278_P2) and the ERC CoG 3D2DPrint (681544).

# References

**\*** of special interest

**\*\*** of outstanding interest

[1] Ayers K: **The potential of proton exchange membrane–based electrolysis technology.** *Current Opinion in Electrochemistry* (2019) **18**: 9-15.

[2] Jensen AW, Sievers GW, Jensen KD, Quinson J, Arminio-Ravelo JA, Brüser V, Arenz M, Escudero-Escribano M: **Self-supported nanostructured iridium-based networks as highly active electrocatalysts for oxygen evolution in acidic media.** *Journal of Materials Chemistry A* (2020) **8**(3):1066-1071.

[3] VahidMohammadi A, Rosen J, Gogotsi Y: **The world of two-dimensional carbides and nitrides (mxenes).** *Science* (2021) **372**(6547):eabf1581.

[4] *Naguib M, Kurtoglu M, Presser V, Lu J, Niu J, Heon M, Hultman L, Gogotsi Y, Barsoum MW: **Two-dimensional nanocrystals produced by exfoliation of $ti_3alc_2$.** *Advanced Materials* (2011) **23**(37):4248-4253.

First paper to report the exfoliation of a MXene from a Max phase. MXene materials could potentially be used in the catalyst layers of OER materials to increase the conductivity of the whole layer.

[5] Alhabeb M, Maleski K, Anasori B, Lelyukh P, Clark L, Sin S, Gogotsi Y: **Guidelines for synthesis and processing of two-dimensional titanium carbide ($ti_3c_2t_x$ mxene).** *Chemistry of Materials* (2017) **29**(18):7633-7644.

[6] Jun B-M, Kim S, Heo J, Park CM, Her N, Jang M, Huang Y, Han J, Yoon Y: **Review of mxenes as new nanomaterials for energy storage/delivery and selected environmental applications.** *Nano Research* (2019) **12**(3):471-487.

[7] He J, Lyu P, Nachtigall P: **New two-dimensional mn-based mxenes with room-temperature ferromagnetism and half-metallicity.** *Journal of Materials Chemistry C* (2016) **4**(47):11143-11149.

[8] Li N, Fan J: **Computational insights into modulating the performance of mxene based electrode materials for rechargeable batteries.** *Nanotechnology* (2021) **32**(25):252001.

[9] Gao G, O'Mullane AP, Du A: **2d mxenes: A new family of promising catalysts for the hydrogen evolution reaction.** *ACS Catalysis* (2017) **7**(1):494-500.

[10] Lim KRG, Handoko AD, Johnson LR, Meng X, Lin M, Subramanian GS, Anasori B, Gogotsi Y, Vojvodic A, Seh ZW: **2h-$mos_2$ on $mo_2ct_x$ mxene nanohybrid for efficient and durable electrocatalytic hydrogen evolution.** *ACS Nano* (2020) **14**(11):16140-16155.

[11]    Jin D, Johnson LR, Raman AS, Ming X, Gao Y, Du F, Wei Y, Chen G, Vojvodic A, Gogotsi Y, Meng X: **Computational screening of 2d ordered double transition-metal carbides (mxenes) as electrocatalysts for hydrogen evolution reaction.** *The Journal of Physical Chemistry C* (2020) **124**(19):10584-10592.

[12]    Browne MP, Sofer Z, Pumera M: **Layered and two dimensional metal oxides for electrochemical energy conversion.** *Energy & Environmental Science* (2019) **12**(1):41-58.

[13]    Tyndall D, Jaskaniec S, Shortall B, Roy A, Gannon L, O'Neill K, Browne MP, Coelho J, McGuinness C, Duesberg GS, Nicolosi V: **Postsynthetic treatment of nickel–iron layered double hydroxides for the optimum catalysis of the oxygen evolution reaction.** *npj 2D Materials and Applications* (2021) **5**(1):73.

[14]    Mayerhöfer B, Speck FD, Hegelheimer M, Bierling M, Abbas D, McLaughlin D, Cherevko S, Thiele S, Peach R: **Electrochemical- and mechanical stability of catalyst layers in anion exchange membrane water electrolysis.** *International Journal of Hydrogen Energy* (2022) **47**(7):4304-4314.

[15]    Lin Z, Shao H, Xu K, Taberna P-L, Simon P: **Mxenes as high-rate electrodes for energy storage.** *Trends in Chemistry* (2020) **2**(7):654-664.

[16]    McAteer D, Godwin IJ, Ling Z, Harvey A, He L, Boland CS, Vega-Mayoral V, Szydłowska B, Rovetta AA, Backes C, Boland JB *et al*: **Liquid exfoliated co(oh)$_2$ nanosheets as low-cost, yet high-performance, catalysts for the oxygen evolution reaction.** *Advanced Energy Materials* (2018) **8**(15):1702965.

[17]    Browne MP, Domínguez C, Colavita PE: **Emerging trends in metal oxide electrocatalysis: Bifunctional oxygen catalysis, synergies and new insights from in situ studies.** *Current Opinion in Electrochemistry* (2018) **7**(208-215).

[18]    Suryanto BHR, Fang T, Cheong S, Tilley RD, Zhao C: **From the inside-out: Leached metal impurities in multiwall carbon nanotubes for purification or electrocatalysis.** *Journal of Materials Chemistry A* (2018) **6**(11):4686-4694.

[19]    Liu A, Liang X, Ren X, Guan W, Gao M, Yang Y, Yang Q, Gao L, Li Y, Ma T: **Recent progress in mxene-based materials: Potential high-performance electrocatalysts.** *Advanced Functional Materials* (2020) **30**(38):2003437.

[20]    Naguib M, Mochalin VN, Barsoum MW, Gogotsi Y: **25th anniversary article: Mxenes: A new family of two-dimensional materials.** *Advanced Materials* (2014) **26**(7):992-1005.

[21]    Browne MP, Dodwell J, Novotny F, Jaśkaniec S, Shearing PR, Nicolosi V, Brett DJL, Pumera M: **Oxygen evolution catalysts under proton exchange membrane conditions in a conventional three electrode cell vs. Electrolyser device: A comparison study and a 3d-printed electrolyser for academic labs.** *Journal of Materials Chemistry A* (2021) **9**(14):9113-9123.

[22]    **Lu Y, Fan D, Chen Z, Xiao W, Cao C, Yang X: **Anchoring $co_3o_4$ nanoparticles on mxene for efficient electrocatalytic oxygen evolution.** *Science Bulletin* (2020) **65**(6):460-466.

*This study along with a few others, have shown that hybrid TMO/MXene materials are superior OER catalysts compared to their counterparts for initial OER activity. However, this study also shows that these TMO/MXene hybrid materials are not stable under prolonged OER studies.*

[23]    **Benchakar M, Bilyk T, Garnero C, Loupias L, Morais C, Pacaud J, Canaff C, Chartier P, Morisset S, Guignard N, Mauchamp V *et al*: **Mxene supported cobalt layered double hydroxide nanocrystals: Facile synthesis route for a synergistic oxygen evolution reaction electrocatalyst.** *Advanced Materials Interfaces* (2019) **6**(23):1901328.

*This study has reported the fabrication of a Co LDH/MXene stable material during intial OER testing. In this study, the edge sites of the MXene flakes are covered with Co LDH catalysts which is attributed to hinder the oxidation process of the MXenes.*

[24]    **Yu M, Zhou S, Wang Z, Zhao J, Qiu J: **Boosting electrocatalytic oxygen evolution by synergistically coupling layered double hydroxide with mxene.** *Nano Energy* (2018) **44**(181-190.

*This paper was one of the first studies to show that TMO and MXene hybrid materials could be alternative OER catalysts to the current state-of-the-art.*

[25]    Shinde PV, Mane P, Chakraborty B, Sekhar Rout C: **Spinel $nife_2o_4$ nanoparticles decorated 2d $ti_3c_2$ mxene sheets for efficient water splitting: Experiments and theories.** *Journal of Colloid and Interface Science* (2021) **602**(232-241.

[26]    Zhu Z, Xu C, Wang Y, Wang L, Chang Z, Fang Z, Liu X, Cheng J: **The high performance nife layered double hydroxides@ $ti_3c_2t_x$/reduced graphene oxide hybrid catalyst for oxygen evolution reaction.** *Journal of Alloys and Compounds* (2022) **894**(

[27]    Cui B, Hu B, Liu J, Wang M, Song Y, Tian K, Zhang Z, He L: **Solution-plasma-assisted bimetallic oxide alloy nanoparticles of pt and pd embedded within two-dimensional $ti_3c_2t_x$ nanosheets as highly active electrocatalysts for overall water splitting.** *ACS Applied Materials & Interfaces* (2018) **10**(28):23858-23873.

[28]    Zhang CJ, Pinilla S, McEvoy N, Cullen CP, Anasori B, Long E, Park S-H, Seral-Ascaso A, Shmeliov A, Krishnan D, Morant C *et al*: **Oxidation stability of colloidal two-dimensional titanium carbides (mxenes).** *Chemistry of Materials* (2017) **29**(11):4848-4856.

[29]    Naguib M, Barsoum MW, Gogotsi Y: **Ten years of progress in the synthesis and development of mxenes.** *Advanced Materials* (2021) **33**(39):2103393.

[30]    Gogotsi Y, Anasori B: **The rise of mxenes.** *ACS Nano* (2019) **13**(8):8491-8494.

[31]     Trotochaud L, Young SL, Ranney JK, Boettcher SW: **Nickel–iron oxyhydroxide oxygen-evolution electrocatalysts: The role of intentional and incidental iron incorporation.** *Journal of the American Chemical Society* (2014) **136**(18):6744-6753.

[32]     Li B, Zhou B, Qu Z, Song Q, Jiang Z: **Theoretical study on fe$_2$c mxene as electrode material for secondary battery.** *Chemical Physics* (2021) **548**(111223.

[33]     Xu D, Stevens MB, Cosby MR, Oener SZ, Smith AM, Enman LJ, Ayers KE, Capuano CB, Renner JN, Danilovic N, Li Y *et al*: **Earth-abundant oxygen electrocatalysts for alkaline anion-exchange-membrane water electrolysis: Effects of catalyst conductivity and comparison with performance in three-electrode cells.** *ACS Catalysis* (2019) **9**(1):7-15.

[34]     Ehelebe K, Escalera-López D, Cherevko S: **Limitations of aqueous model systems in the stability assessment of electrocatalysts for oxygen reactions in fuel cell and electrolyzers.** *Current Opinion in Electrochemistry* (2021) **29**(100832.